\documentclass[12pt,titlepage]{article}

\usepackage{latexsym}
\usepackage{amsfonts}
\usepackage{amssymb}
\usepackage{epsfig}

\begin{document}
%
\title{Global Characteristics of Nucleus-Nucleus \\
       Collisions in an Ultrarelativistic Domain}

\author{M.V.~Savina, \underline{S.V.~Shmatov}\thanks{e-mail: shmatov@lhe.jinr.ru},
    N.V.~Slavin, P.I.~Zarubin\\
   \it Joint Institute for Nuclear Research, Dubna, Russia}
\maketitle

\begin{abstract}\
The global observable distributions of nucleus-nucleus collisions at high energy
are studied. It is shown that these distributions are sensitive to
interaction dynamics and can be used to investigate the evolution of dense nuclear
matter. At the LHC energy scale ($\sqrt{s_{NN}}=5$ TeV in the heavy ion mode),
a central bump over pseudorapidity plateau in the differential
distributions of total transverse energy flows induced by the jet quenching
effect was obtained. The energy-loss dependence on the energy for various colliding
ion species was explored.

Also a new scheme for collision geometry determination is
proposed as well. It is based on the correlation between
transverse energy flow in the pseudorapidity interval $3<|\eta|<5$ and an
impact parameter.
\\[2mm]
\end{abstract}


%


\section{Introduction}
Global characteristics of nucleus-nucleus collisions available in the
future experiments at high energy (the LHC energy scale) are differential distributions
of  total transverse energy flows, $dE_T/d\eta$, as well as  electromagnetic
energy and charged
multiplicity ones, $dE_{\gamma}/d\eta$  and $dN_{ch}/d\eta$.
Measurements of the global distributions allows one to establish  quite
general rules of nucleus-nucleus  collision dynamics
up to the highest collision energy frontier over a widest rapidity range and
verify some critical predictions of quark-gluon plasma formation models in a
sufficiently simple way.

In addition to the general physics interest the global event response provides
an estimate of a collision impact parameter for correlated studies in  specific
reaction channels like jet and $W$, $Z^0$ production.

\section{Jet Quenching Manifestation in Global Energy Flows}

The dense  matter formation (or quark-gluon plasma) in relativistic nuclear
collisions  is predicted by many models \cite{Satz85}.
One of the discussed features of such a state of a nuclear matter is  energy
losses of scattered partons in final state interactions with a dense nuclear matter called
jet quenching \cite{Gyulassy90, Gyulassy91}. Among other effects originated by a jet quenching one may expect a significant
modification in the differential distributions mentioned above.  Indeed  an
indication is found for ultrarelativistic energy domain on  the appearance of a wide bump
in the  interval $-2<\eta<2$ over a pseudorapidity plateau of such distributions
due to jet quenching \cite{Savina98}.
The bump existence problem   rises a  new  question of principle  --
whether the asymptotic behaviour of the distributions established already at
the lower energy scale will be  broken in a new energy domain, or not. A calorimeter
with a wide enough acceptance ($-5<\eta<5$) gives a chance to obtain a definite answer
to the question.

Our calculations has been done in the framework of the HIJING model \cite{Wang97}
for the energy range from 1 up to 5 TeV/nucleon. In accordance to the HIJING already
$\approx$80\% of the total transverse energy flow are calculated by a perturbative QCD
application. A minijet production (i. e. jets with $p_t \ge$ 2 GeV)
becomes a dominant feature of multiple processes in the considered energy region.
Soft hadron-hadron processes are
simulated as classical strings with kinks and valence
quark ends following the FRITIOF model \cite{Anderson87,Nilson87}. The number of jets in
inelastic nucleon-nucleon collisions is calculated in the eikonal approximation. The model
provides the dependence of the parton structure function on a collision
impact parameter.

The most important feature of the HIJING model is
the energy losses $dE/dx$ of partons traversing a dense
nuclear matter.  High energy quark and gluon losses in a hot
chromodynamic matter are estimated in \cite{Bjorken82,Thoma91}.
It was shown that the dominant
mechanism of energy losses is a radiative one due to a gluon emission
inside a narrow jet cone (Bethe-Heitler limit) in soft final state interactions
(bremsstrahlung) in accordance with the relation used in the HIJING:

\begin{equation}
  \frac{dE}{dx} \approx
  \frac{C_{F}\alpha_s}{\pi} \mu^2 ln\frac{3ET}{2\mu^2}
  (ln\frac{9E}{\pi^3T}+\frac{3\pi^2\alpha_s}{2\mu^2}T^2)
\label{eq:1}
\end{equation}
where $E$ is the propagating parton energy, ${\mu_D}^{-1} \sim 1/gT$ the screening constant
(Debye cutoff scale), $\alpha_s$ the running coupling constant.
In the HIJING model
this mechanism is fulfilled as consecutive transmissions of  parton energy fractions
 from one string configuration to
another.

It is supposed that energy losses are
proportional to a distance $l$ passed by a jet after a last interaction
and occur only in the transverse direction within the nucleus radius.
An interaction proceeds until a parton jet stays inside the considered
volume and the jet energy exceeds a jet production energy limit.
The space-time picture of the evolution of a parton shower  is not
considered in the HIJING and  the dense matter formation is
introduced by a phenomenological way.

\begin{figure}
\begin{center}
\epsfig{figure=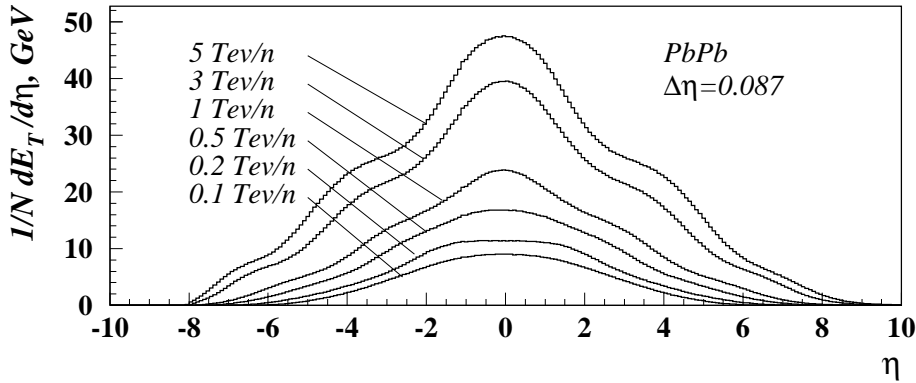, height=5cm}
\end{center}
\caption{Differential distribution of total transverse energy $dE_T/d\eta$
(GeV) over pseudorapidity $\eta$ for 10000 minimum bias
PbPb collisions at $\sqrt{s_{NN}}$= 5, 3, 1, 0.5, 0.2, 0.1
TeV/nucleon. Normalized per number of events.}
\label{fig:1}
\end{figure}

In fig.\ref{fig:1} differential
transverse energy distributions $dE_T/d{\eta}$ are presented for minimum bias Pb-Pb
collision at the CMS energy $\sqrt{s_{NN}}$ equal to 5, 3, 1, 0.5, 0.2, 0.1 TeV/nucleon.
A  distinctive feature of the distributions is the appearance  of a
central pseudorapidity bump (-2$<\eta<$2).

  One may conclude that the bump becomes distinguishable over a plateau starting
from an energy value larger than 3 TeV/nucleon.  For lower energy
values this effect is not so profound due to smaller value of parton energy losses
in the nuclei. This fact induced by decreasing the Debye
screening constant of nuclear matter at a lower energy.

\begin{figure}
\begin{center}
\epsfig{figure=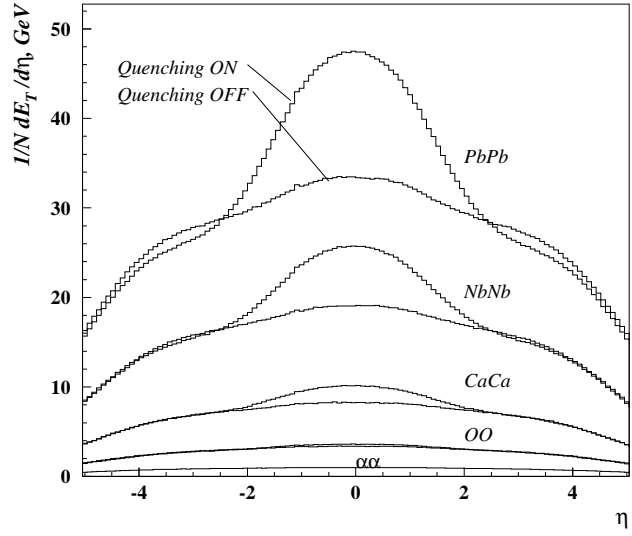, height=7cm}
\end{center}
\caption{Differential distribution of total transverse energy flow $dE_T/d\eta$
(GeV) over pseudorapidity $\eta$ for 10000 minimum bias
Pb-Pb, Nb-Nb, Ca-Ca, O-O, $\alpha-\alpha$ collisions at
$\sqrt{s_{NN}}$=5 TeV/nucleon  with and without jet quenching. Normalized per number of
events; $\eta$  bin size is 0.087.}
\label{fig:2}
\end{figure}

We followed a jet quenching sensitivity to a mass number for
lighter colliding nuclei. Energy losses of hard parton jets depend
on a parton path in a dense matter like $\Delta E=ldE/dx$.
Therefore reduction of the colliding nucleus radii might lead to a bump reduction. This
provides an additional verification of the jet quenching effect \cite{Savina981}.  Fig.\ref{fig:2} shows that
lead-lead collisions demonstrate maximum quenching dependence  while in lighter
ion cases the bump becomes less and less profound. It's interesting to note
that the shadowing induced modification  of the parton distribution
 functions can be estimated by means of
absolute value measurements of the transverse energy flow density for various ion species.

\section{Impact Parameter Estimation}

A secondary interaction effect as a jet quenching
modifies only the central rapidity part leaving both fragmentation regions ($3\le|\eta|\le5$)
practically unchanged. This circumstance can be used for estimation of a collision impact
parameter.
The average number of minijets produced in
nucleus-nucleus collisions in the pseudorapidity interval ${\Delta \eta}$ is related with
a collision impact parameter $b$:

\begin{equation}
   \bar N_{AA}(b,\sqrt{s},p_0)_{\Delta \eta}=T_{AA}(b)\sigma_{jet}(\sqrt{s},p_0)_{\Delta \eta}
\label{eq:2}
\end{equation}
where $T_{AA}(b)$ is the nuclear density overlap function of two colliding
nuclei calculated with the  assumption of the
 Wood-Saxon nuclear density distribution   $\rho(r)$,
$\sigma_{jet}$ the minijet production  cross-section.

An average transverse energy flow in $3\le|\eta|\le5$
intervals is related with a collision impact parameter as

\begin{equation}
  \bar E_T(b,\sqrt{s},p_0,3\le|\eta|\le5)=T_{AA}(b)
  \sigma_{jet}(\sqrt{s},p_0,3\le|\eta|\le5)<E_t>_{\Delta \eta}
\label{eq:3}
\end{equation}
where $E_t$ is the mean transverse energy per a minijet. The experimental data shows
that parton structure functions using to calculated $\sigma_{jet}(\sqrt{s},p_0,3\le|\eta|\le5)
<E_t>_{\Delta \eta}$ stay relevant until transverse energy per minijet $E_t\le$300 GeV.
In the considered pseudorapidity range the minijet energy $E_t$ is of the order of
10 GeV. Thus, in this way  the nuclear collision geometry (i.e.  impact parameter) can well
be defined by means of the wide acceptance calorimeter with minimal dependence
on  collision dynamics details in the central region.

\begin{figure}
\begin{center}
\epsfig{figure=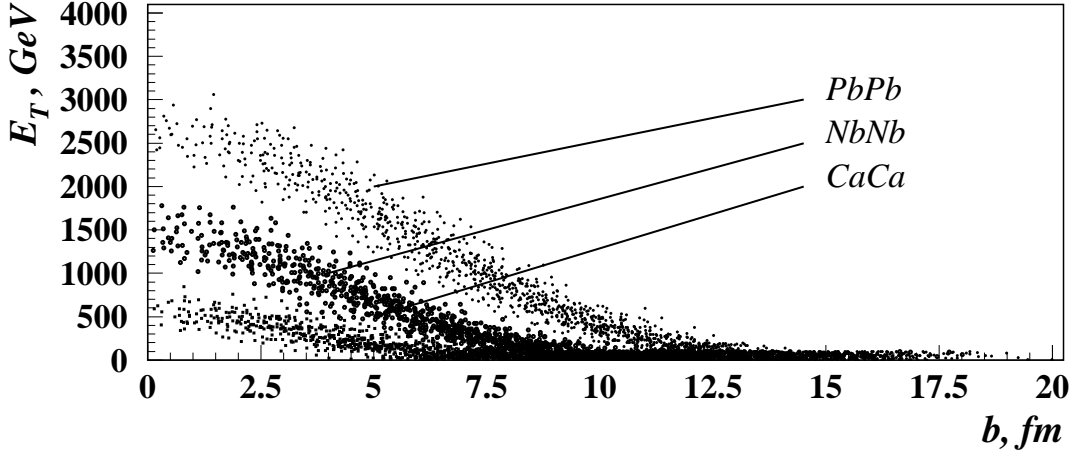, height=6cm}
\end{center}
\caption{Correlation between transverse energy flow per collision $E_T$ (GeV) in the
pseudorapidity direction (3$\le |\eta| \le $5) and collision impact parameter $b$ (fm). From top
to bottom: PbPb, NbNb, CaCa collisions at $\sqrt{s_{NN}}$=5 TeV/nucleon.}
\label{fig:3}
\end{figure}

Fig.\ref{fig:3} shows correlations between $E_T$ and $b$ for CaCa,
NbNb, PbPb collisions allowing to conclude that an impact
parameter can be estimated with 1.5 fm precision.








\end{document}